# Informatics Issues
# Used in the Production Dashboard

**Alin Isac, Claudia Isac**
**University of Petroşani, România**

**ABSTRACT**. The aim of this paper is to present some computer aspects regarding the implementation and the employing of a dashboard in relation to the production activity. The paper begins with the theoretical presentation of the managerial perspective regarding the necessity of using the dashboard. The main functions of the dashboard in the production activity and the way it is employed are presented in the second part of the paper.

## 1. Dashboard General Characteristics

The one of the most effective ways of reflecting the process of improvement is management dashboard that allows increased flexibility through the removal of unfavourable deviations recorded in the functional parameters and indicated that while the anchor point of improvement tools piloting used by managers.

Although the literature is using few concepts on the dashboard as exposed in [Eke05] [Nic00] [Ver01] [ZBC96], it is defined as a set of current indicators and presented in a synoptic, default, on the main results of the work envisaged and factors that make them run efficiently. From its intended use, namely to allow the manager of a company or a part thereof, to ascertain the progress of its work and assigned to the systemic point of view of methodology, the dashboard can be defined as a group coherent rules, methods, procedures and decision-making situations through which information shall forecasting, monitoring and evaluation of business or organizational components of the company as a whole.

Dashboard can be designed and built at any level of management or any manager, but there is a dashboard type, valid for all managerial functions. While paintings by board type is not extensive, use this tool enables management of the classification after several criteria.

Although a variety of known ways of effective implementation, the dashboard is effective if it meets the four main functions which define the content, namely: the warning on the adverse circumstances, the assessment-diagnostic results, the information manager state of the managed domain, the decision to allow the rapid adoption of decisions by managers.

Because functions that meets and variables that is, the dashboard allows the separation of anticipated findings on the effects and relationship management processes and thus alerting managers on indicators that have reached a critical area. Thus, it outlines very important position of the Dashboard in all ways to increase the effectiveness of work managers and systems management.

Advantages of using the Dashboard shows the need to extend the conception and realization of the Dashboard to managers located in all hierarchical levels and in all fields of activity. Diversity of benefits arising from implementation of the Dashboard, shows the following: that the best decision by making available to managers of information consistent, rigorous, aggregates, expressive and accessible on the main aspects of the firm or in the lead; approach information relating to management in a systemic vision that allows managers to efficiently allocate the human, material, financial, information and time, making better use of working time of managers by targeting their activities towards the key issues facing the company ; increase operability adapt to the subsystem operational factors with disturbances facility extending the use of electronic equipment for information processing.

## 2. Dashboard TI Component in Management System

They emphasize the system of management of the Dashboard - component subsystem methodological - and highlights interdependent other subsystems management: information, organization and decision-making. Thus, models information in the Dashboard compiled for the operational and organizational are transmitted through the subsystem information (especially the science of it) to subsystem making the results of basic auxiliary or service in the form of indicators or signals warning as shown in Figure 1. If the values they presented irregularities important parameters prescribed managers who have processed the information through the dashboard, will take appropriate decisions.

Starting from the relationship between subsystems, the dashboard can be built in a hierarchical design to include information corresponding to each level of activity. For example, a company in the energy industry, a **dashboard** may include first a marketing dashboard, in turn, consists of paintings on its customers, by region, the transport networks and on-board paintings produced on sections or workshops (Department of turbines, boilers Department, household fuel, etc.) that can be incorporated into a dashboard of productive activity level power and financial dashboard that highlights, in particular, indicators of liquidity and solvency of the company.

In industrial firms can develop a dashboard-level production system which will monitor progress in real time operation of productive activities. To implement this system of management in the operational management of production to create a system of integrated management of production activities based on a system of data entry and by referencing all "agendas" of management in a coherent whole.

Order to implement the Dashboard of production is to give managers information about the functioning of operational blocks in real time (by presenting the most important evolution parameters), information about stage of production to achieve within one hour (average values of the parameters during a hour) and average track parameters on different time periods [Jab02]. Means of realization of this picture are computer equipment and software specifically assimilated to retrieve and process data, and their choice of technical and quality should consider the structure of the company, the dispersion territorial operational groups, the degree of saturation of channel information, the possibility of simultaneous transmissions of data. Ensure the functionality of the Dashboard production is done using the terminal compartment in functional groups on the operational mode to permit the application of interactive activities and provide assistance during the production implementation.

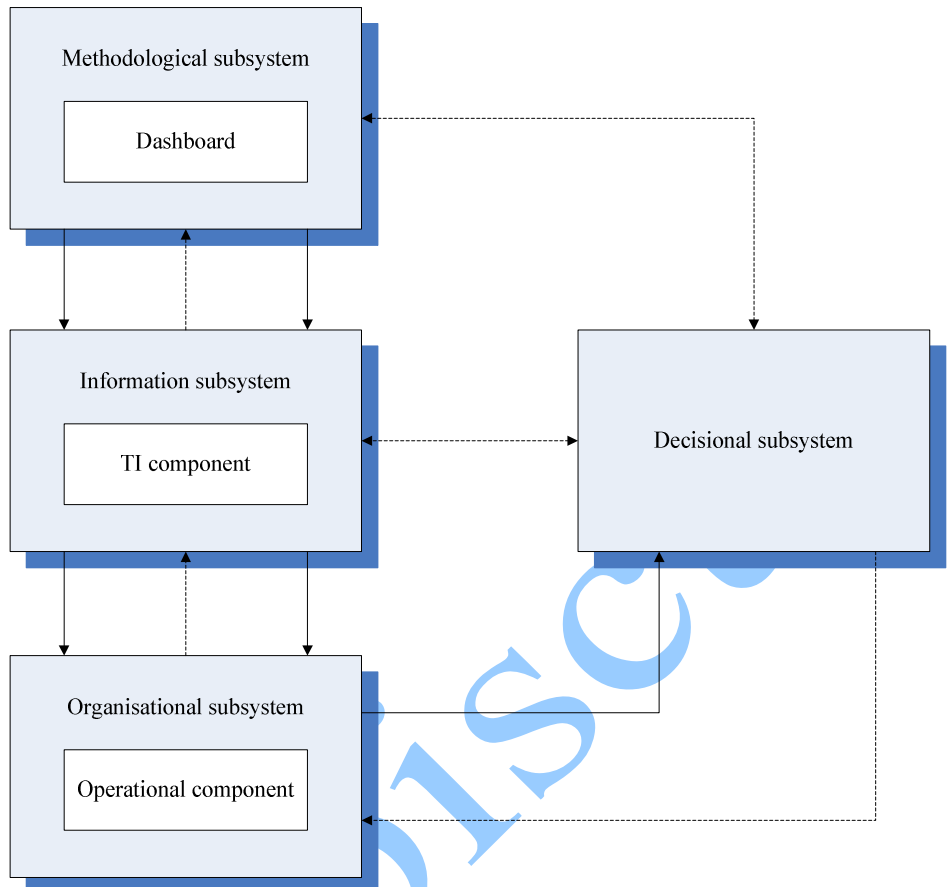

**Figure 1 Relationship between panel subsystem and other subsystems methodological management**

## 3. Production Dashboard TI Functions

Dashboard tracking manufacturing activity has the following functions:

A. Accessing the Dashboard. User the Dashboard is available on the monitor screen, an icon that allows direct entry application. By accessing the application starts running, viewing from the main menu. The three types of functions of the Dashboard can be accessed directly from the main menu (Figure 2) by selecting the desired function, or instantaneous values for the main technological parameters, the values of these different time intervals and functions of monitoring parameters.

160

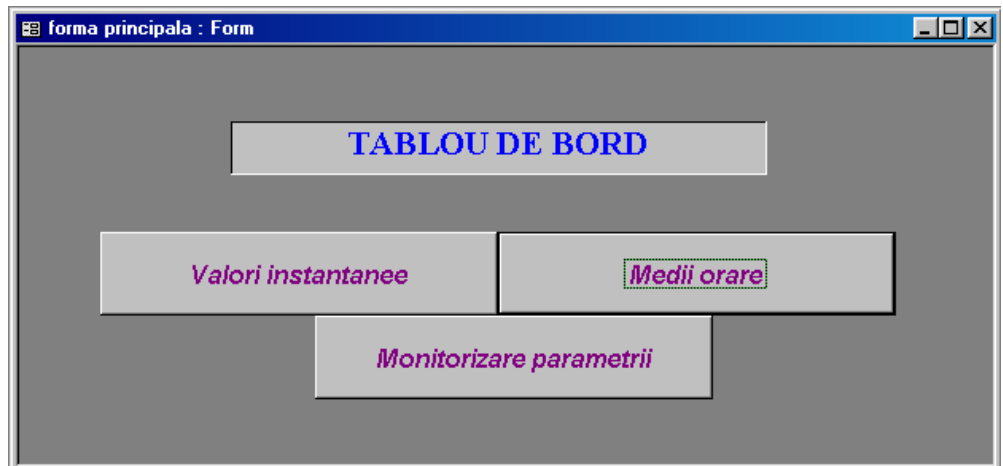

**Figure 2 Main menu of the Dashboard**

B. The function of real-time view of the evolution of power plant technological parameters. Through periodic processing of the data contained in the "band" of production will obtain summary information on the dynamics of productive achievement. In general, the panel used in functional compartments provides summary information on the use of facilities and equipment technology and production highlights the divisions and daily and cumulative daily deviations from pre-production and causes of such deviations. Since the production is influenced by a large number of parameters through the dashboard, the manager seeks the appointment of the best indicators of the state system and its main task prompt intervention to adjust the production process. If discrepancies arise, caused by disturbances in the system, will make decisions accordingly to remove them. These parameters are selected through a database containing a large number of records that are archived by category of terminals where they were measured. The portfolio of these parameters may vary depending on the priorities that are raised by the manager using the Dashboard, in the sense that it may waive part of the parameters on which it considers relevant for the substantiation of decisions or increase their number or on a single block, or all, particularly where equipment and aggregates whose duration of operation is lower and shows a higher damage than the average.

C. Monitoring Function. Through this function, acting manager of the dashboard, usually the deputy technical director of production, are able to calculate absolute or relative values of parameters by changing them so as to obtain the optimal functioning of all operational and functional blocks in time monitoring. Thus, the dashboard allows the representation of on-line evolution following parameters: the calculation of specific prescribing the possibility functioning; parameters presenting the archive database or processed, etc.

D. Delimitation of the analysis. For the analysis parameters for a given period will be viewed over the selected user. He enrols the period in the dashboard or in fields in the top frame of the shape and the supervised production department, after which the user uses the option of choice for the parameters you want graphics.

**Conclusions**

Important changes taking place in the company's internal and external cause managers to ensure continued improvement of the management system and subsystem default methodology using systems, methods and management techniques more efficient. Ensure a subsystem methodological functional and effective may be the empirical fact that improving the subsystem management is an approach developed the type of professional who have a permanent character and to ensure efficiency of business management.